\documentstyle[12pt]{article}

\newcommand{\nc}{\newcommand}
\nc{\beq}{\begin{equation}}
\nc{\eeq}{\end{equation}}
\nc{\beqa}{\begin{eqnarray}}
\nc{\eeqa}{\end{eqnarray}}

\def\gsim{\mathrel{\rlap{\lower4pt\hbox{\hskip1pt$\sim$}}
    \raise1pt\hbox{$>$}}}       

\input epsf
\newwrite\ffile\global\newcount\figno \global\figno=1

\def\writedef#1{}
\def\figin{\epsfcheck\figin}\def\figins{\epsfcheck\figins}
\def\epsfcheck{\ifx\epsfbox\UnDeFiNeD
\message{(NO epsf.tex, FIGURES WILL BE IGNORED)}
\gdef\figin##1{\vskip2in}\gdef\figins##1{\hskip.5in}
\else\message{(FIGURES WILL BE INCLUDED)}%
\gdef\figin##1{##1}\gdef\figins##1{##1}\fi}
\def\figinsert{}
\def\ifig#1#2#3{\xdef#1{fig.~\the\figno}
\writedef{#1\leftbracket fig.\noexpand~\the\figno}%
\figinsert\figin{\centerline{#3}}\medskip\centerline{\vbox{\baselineskip12pt
\advance\hsize by -1truein\center\footnotesize{  Fig.~\the\figno.} #2}}
\bigskip\endinsert\global\advance\figno by1}
\def\endinsert{}

\begin{document}



\title{\large \bf Towards a High Energy Theory for the Higgs Phase of
Gravity}

\author{\normalsize Michael L. Graesser$^a$, Ian Low$^b$, and Mark B.
Wise$^a$ \\
\normalsize \it $^a$California Institute of Technology, 452-48,
Pasadena, CA 91125 \\
\normalsize \it $^b$School of Natural Sciences, Institute for Advanced
Study, Princeton, NJ 08540}

\date{}

\maketitle

\vspace{-7cm}

\hfill{\small CALT-68-2575}

\vspace{6cm}
\begin{abstract}
Spontaneous Lorentz violation due to a time-dependent expectation
value for a massless scalar has been suggested as a method for
dynamically generating dark energy. A natural candidate for the
scalar is a Goldstone boson arising from the spontaneous breaking
of a $U(1)$ symmetry. We investigate the
low-energy effective action for such a Goldstone boson in a
general class of models involving only scalars, proving that if the
scalars have standard kinetic terms then at the {\em classical}
level the effective action does not have the required features for
spontaneous Lorentz violation to occur asymptotically $(t
\rightarrow \infty)$ in an expanding FRW universe. Then we study
the large $N$ limit of a renormalizable  field theory with a
complex scalar coupled to massive fermions. In this model an
effective action for the Goldstone boson with the properties
required for spontaneous Lorentz violation can be generated. 
Although the model has shortcomings, we feel it represents 
progress towards finding a high energy completion for 
the Higgs phase of gravity. 
\end{abstract}


%
\setcounter{footnote}{0} \setcounter{page}{1}
\setcounter{section}{0} \setcounter{subsection}{0}
\setcounter{subsubsection}{0}

\newpage

\section{Introduction}
 For a long time physicists hoped that the value of the
cosmological constant would be zero, since that might be easier to
understand than a small but a nonzero value. But the inference of
the existence of dark energy from the supernova data, and its
concordance with the Cosmic Microwave background, gravitational
lensing and large-scale structure data, continue to suggest
otherwise. These observations give support to the 1987 prediction
of Weinberg's that, all else being equal, the cosmological
constant cannot be orders of magnitude larger than the cold dark
matter component, otherwise galaxies would not have formed
\cite{weinberg}. The discovery of only a single new particle at
the Large Hadron Collider - the Higgs boson - would also give support to the logical
possibility of a landscape \cite{landscape} in which some of the
otherwise seemingly unrelated parameters of the Standard Model are
strongly correlated in order that atoms and galaxies can exist
\cite{anthropic}.

An alternate - and perhaps
testable - possibility is that there is a dynamical explanation for
why the cosmological constant is zero. The mechanism
responsible for canceling the expected large contributions
from Standard Model particles could perhaps be tested
by performing measurements in our own universe.
But then there are two
remarkable things to explain: what is the dark energy; and
why it began to dominate the evolution of the universe only recently.
An elegant explanation is provided by tracker \cite{tracker} 
versions of quintessence \cite{quintessence}. 
However, they tend to give 
the wrong equation of state for the dark energy. Another interesting 
possibility is that the energy scale of the dark energy is 
tied to the neutrino mass \cite{massvarnu}.

The work 
of \cite{weinberg2} rules out an explanation for why the cosmological 
constant is zero, where the 
vacuum expectation value of a field relaxes to a value that 
cancels the ``bare'' cosmological 
constant. It has been suggested that topology-changing 
configurations in Euclidean quantum gravity force 
the cosmological constant to be zero \cite{coleman2}. 
See however, \cite{polchinski2}. 

The `ghost condensation' or
`Higgs phase of gravity' proposal \cite{nima1}
provides a non-trivial
scenario in which the dark energy arises dynamically
from a time-dependent scalar field. There has been 
considerable recent interest in this proposal. See for 
example \cite{recentpapers}. 

The starting point for ghost condensation 
is the existence of a scalar with a shift
symmetry
\beq
\phi \rightarrow \phi + \alpha
\eeq
that guarantees that the effective action for $\phi$ involves only
derivatives acting on $\phi$:
\beq
S= \int d^4 x {\cal L}(\partial _{\mu} \phi)~.
\eeq
Such a theory admits solutions of the form
\beq
\phi= c t 
\label{asolution0}
\eeq
where $c$ is a constant.
For a free theory in an expanding universe, these
solutions become irrelevant at late times. In a non-trivial
theory though, they can be important \cite{nima1}.

\begin{figure}[t!]
  \centerline{\mbox{\epsfysize=5truecm
\hbox{\epsfbox{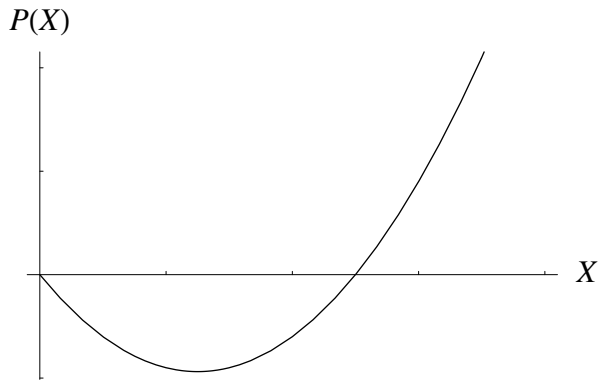}} } }
  {\caption[1]{Non-trivial effective action with
asymptotically Lorentz-violating solution.}
\label{fig:figgc1} }
\end{figure}

The effective action that generates such solutions
has the form
\beq
{\cal L } = P(X)
\eeq
where $X \equiv (\partial _{\mu} \phi)^2$ and
we have ignored any operators involving two or more
derivatives on $\phi$, since they do not contribute
to the equations of motion for solutions of the form
(\ref{asolution0}).
However, depending on the
form of $P$, in an expanding
universe not all of these solutions for $X$ will remain
static. The equation of motion for $X$ is
\beq
\partial ^{\mu} \left(a ^3 P^{\prime}(X) \partial _{\mu} \phi \right)
=0
\eeq
where $a(t)$ is the scale factor.
Focusing on solutions that depend only on time,
\beq
P^{\prime}(X) \partial_0 \phi = \frac{\tilde{c}}{a^3} ~.
\eeq
The subsequent
evolution of $X$ depends on the form of $P$. If there is a
nearby
point $X_*$ where
\beq
P^{\prime}(X_*)=0~,
\eeq
then $X$ will be driven to $X_*$ \cite{nima1}.
Such points with $P^{\prime \prime}(X_*)>0$ have
the correct sign for the two-time derivative
term in the Lagrangian for
small fluctuations in $\phi$. Furthermore,
the operator $(\Box \phi)^2$ must be present with a negative
coefficient to avoid any spatial instabilities \cite{nima1}.
If $P(X_*)$ is negative, as in Figure \ref{fig:figgc1},
then
the dark energy is
positive. By contrast,
if $P^{\prime}$ has no zeros, say $P \sim X^n$ for some
range of $X$, then in that interval $P(X) \sim a^{-6n/(2n-1)}$
and $P(X)$ becomes increasingly irrelevant at late times.

\begin{figure}[t!]
  \centerline{\mbox{\epsfysize=5truecm
\hbox{\epsfbox{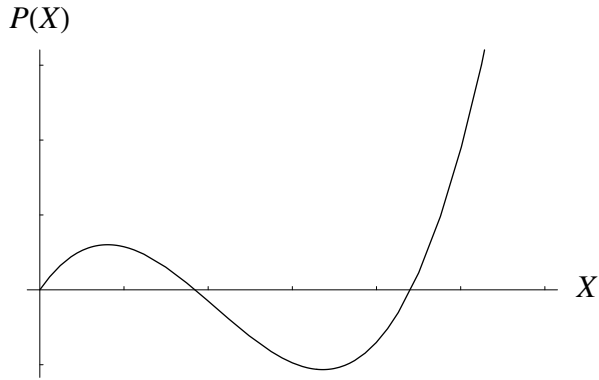}} } }
  {\caption[1]{Another possibility for
a non-trivial effective action with
an asymptotically Lorentz-violating solution, but no ghost near
the origin.}
\label{fig:figgc2} }
\end{figure}

In the example provided in Figure \ref{fig:figgc1},
the slope near the origin is negative so
the theory has a ghost. One might be concerned whether
such a ghost could be generated in a theory that
at a fundamental level does not have one. It is also possible that
there exists an $X_*$ where $P^{\prime}(X_*)=0$ but that near the
origin the slope is positive. An example
of such a $P$ is provided in Figure \ref{fig:figgc2}.

In this paper we investigate whether a renormalizable field theory
with a spontaneously broken $U(1)$ symmetry can generate
a $P$ of either form for the Goldstone boson, $\phi$. In particular,
we will
ask whether there can exist solutions with $P^{\prime}=0$ when the
underlying field theory is taken to have standard kinetic terms for
its fields. 

We first consider integrating out the heavy degrees of freedom at the
classical level and give a short proof that the spontaneous breaking
of a global $U(1)$
symmetry in
a general theory of interacting massive
scalars does not lead
to a $P(X)$ with any extrema.
Then we go on to consider a case where the complex scalar, whose
vacuum expectation value spontaneously breaks the $U(1)$ symmetry,
couples to $2N$ massive fermions. We integrate out the fermions in the
large $N$ limit and find that a $P(X)$ for the Goldstone boson like
that shown in Figure \ref{fig:figgc1} can be generated at strong coupling.
Even though the theories we are considering are renormalizable, 
they do require 
regularization, and hence a cutoff. 
The model we constructed 
does not provide a conventional high energy theory for the Higgs phase of
gravity because the fermion masses are necessarily near the
ultraviolet cutoff of the full theory. Furthermore the low energy 
theory for the Goldstone boson 
is unacceptable because it contains 
a spatial instability. Despite these shortcomings, 
our work shows that a $P(X)$ like that in 
Figure \ref{fig:figgc1} can be generated from an underlying 
theory with normal kinetic terms.

\section{Classical Physics}
\label{classicalphysics}

To start with consider the very simple example of a single complex
field $\Phi$ whose vacuum expectation value (vev)
spontaneously breaks a
$U(1)$ global symmetry. The
renormalizable Lagrangian is
\beq
{\cal L}_0 = \partial _{\mu} \Phi ^* \partial ^{\mu} \Phi
- \frac{\lambda}{4}(|\Phi|^2 - v^2)^2 ~.
\eeq
It is convenient to parameterize the fields as
\beq
\Phi = \left(v + \frac{b}{\sqrt{2}} \right) e^{i {\phi}} ,
\eeq
where $\phi$ is the Goldstone boson, chosen to not be
canonically normalized.
$b$ is a massive field
that we will integrate out of the theory at the classical level
to generate an effective action for the Goldstone boson.

The theory has both a $U(1)$ global symmetry - here realized linearly
- and
a time translation invariance. The solution
\beq
\phi= c t
\label{asolution}
\eeq
preserves a linear combination of the time translation
symmetry and the non-linearly realized shift symmetry
\cite{nima1}. We are seeking solutions to the exact equations of
motion which respect that symmetry.
Inspecting $\Phi$, we note that the magnitude of
$\Phi$ though, is invariant only if $b$ is static.
This means that we are looking for solutions to the
exact classical equations of motion in which
$b$ is static and $\phi$ is given by (\ref{asolution}).

It is further convenient to define $y \equiv \sqrt{2} b v
+\frac{1}{2} b^2$.
Note that $|\Phi|^2=v^2+y$ and $y \geq -v^2$. $y=-v^2$ is
a singular point since there the $U(1)$ is unbroken.
In terms of these variables
\beq
{\cal L} = \left[\frac{1}{2}\frac{(\partial _{\mu } y )^2}{v^2+y}
+ {X}(v^2+y) - \frac{\lambda}{4} y^2  \right]~.
\eeq
Next we integrate out $y$ classically, which amounts to solving its
classical equations of motion in terms of $X$. As argued
above, we assume that there is no time variation in $b$.
The solution is then
\beq
y=2 \frac{X}{\lambda}~.
\eeq
Substituting this solution for $y$ back into the action gives
the effective action for $X$:
\beqa
P(X) &=& X \left(v^2+2\frac{X}{\lambda}\right)
- \frac{1}{ \lambda} X^2
  \nonumber \\
&=& X\left(v^2 + \frac{X}{\lambda }\right)~.
\eeqa
The effective action for $X$ is monotonic and has no extrema.
In particular,
\beq
P^{\prime}=v^2+2\frac{X}{\lambda }=(v^2+y[X])=\Phi^* \Phi ~.
\label{example1P}
\eeq
Notice that $P^{\prime}(X)$ is positive for all $X$. We now proceed
to show that at the classical level this will always be the case.

Consider a general $U(1)$ theory with $N$ scalar fields $\chi_i$
each of charge $q_i$ (some of which may be zero) and
\beq
{\cal L }
= K- V ~.
\eeq
Here $V$ is the most general potential consistent
with the global symmetry (it does not have to be renormalizable.). $K$
is the standard kinetic terms for the scalar fields
\beq
K=\sum_ i \partial_{\mu}  \chi^* _i \partial^{\mu} \chi_i~.
\eeq
For any value of each $\chi_i$, there is a direction in field
space that keeps $V$ constant. Promoting this direction, given by
\beq
\chi_i \rightarrow \chi_i e^{i q_i \phi} ~,
\eeq
to a field $\phi$, identifies the Goldstone boson.

To obtain the effective
action for $\phi$, we look for solutions that preserve the
unbroken combination of time translation and $U(1)$ symmetries, so
\beq
\chi_i=f_i(X) e^{i q_i \phi} ~.
\eeq
Recall, $X \equiv (\partial _{\mu}\phi)^2$. $f_i$ is static, but may
be complex
and depend on $X$. The effective action for $f_i$ and $X$ is
\beq
L= \sum_i q^2_i |f_i|^2 X - V[\{f_j\}]~.
\eeq
The equation of motion for $f^*_i$ gives
\beq
q^2_i X f_i = \frac{\partial V}{\partial f^*_i} ~.
\eeq
Solving the $N$ equations determines $f_i(X)$. The
effective action for $X$ is then
\beq
P(X) = \sum_i q^2_i X |f_i(X)|^2 - V[\{f_j(X)\}]
\eeq
so that
\beqa
P^{\prime}(X) &=& \sum_i q^2_i |f_i|^2 +\sum_i q^2_i X
\left[ f^*_i \frac{\partial f_i}{\partial X} +\hbox{h.c.} \right]
-\sum_i \left[\frac{\partial V}{\partial f_i} \frac{\partial
f_i}{\partial X}
+\hbox{h.c.} \right] \nonumber \\
&=& \sum_i q^2_i |f_i|^2 \geq 0~.
\label{generalformula}
\eeqa
In this class of theories, $P$ is monotonically increasing.
From this expression
we see that
$P$ can have an extrema only if all the fields vanish:
$f_i=0$. At this location the $U(1)$ global symmetry is unbroken and
there is no Goldstone boson. Note that,
\beq
P^{\prime}(X)= \left.\frac{\partial K[\{f_i\},X]}{\partial X}\right|  _{f_i
=f_i(X)}~.
\label{Keqn}
\eeq
(\ref{Keqn})
is also valid for any $K$ that is analytic in $\partial_{\mu} \chi^*_i
\partial^{\mu} \chi_i$.

Eq. (\ref{generalformula}) reproduces the previous
result for $P^{\prime}$ in the one field model considered earlier,
where, $q=1$ and
$|f|^2 =|\Phi|^2= v^2+y$.

\section{Towards A High Energy Completion}
\label{toymodel}

The toy model we consider
has a complex scalar $\Phi$
and two Dirac multiplets of fermions $\psi_1$ and $\psi_2$.
Under the $U(1)$ global symmetry, the scalar has charge
$+1$ and the fermions have charge $+1$ and $+2$ respectively.
The model also has a global $SU(N)$
symmetry, where the fermions each transform under the fundamental
representation and the scalar is neutral. We suppress the $SU(N)$ indices
on the fermions.

The most general renormalizable Lagrangian, consistent
with these symmetries, is
\beqa
{\cal L} & = & \partial _{\mu} \Phi ^* \partial ^{\mu} \Phi
-\frac{\lambda}{4} ( |\Phi|^2-v^2)^2 \nonumber \\
 & & +  \sum _{i=1,2}
\left( i\bar{\psi}_i \gamma^{\mu} \partial_{\mu}
\psi_i -  m_{\psi} \bar{\psi}_i \psi_i \right)
- g \Phi \bar{\psi}_2 \psi_1 -g \Phi^* \bar{\psi}_1 \psi_2~.
\label{toymodelL}
\eeqa
Note that all the fields have the conventional sign for
their kinetic terms. We have performed a field redefinition to
make $g$ real and for simplicity have taken the two types
of fermions to have equal masses $m_{\psi}$.

\begin{figure}[t!]
  \centerline{\mbox{\epsfysize=5truecm
\hbox{\epsfbox{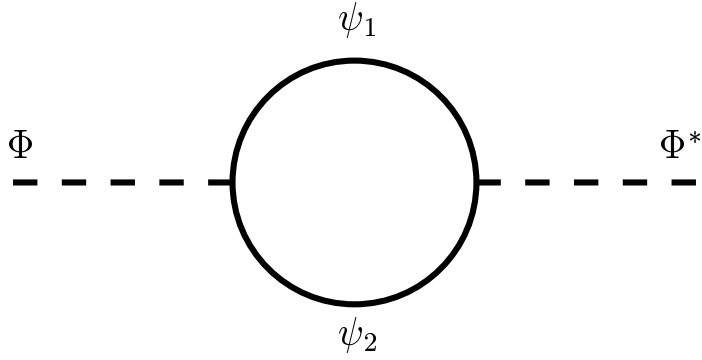}} } }
  {\caption[1]{Leading one-loop diagram contributing to effective 
action for $\Phi$.}
\label{fig:oneloop} }
\end{figure}

Loops involving the fermions and scalar
self-interactions generate an effective action for
$\Phi$. To generate a ghost-like kinetic term for $\Phi$,
quantum corrections must overcome the tree-level kinetic
term. This implies that perturbation theory
is not valid. Hence we need the exact
effective action
to conclude that quantum
corrections have flipped the sign of the kinetic
term for $\Phi$. The large quantum corrections
come from fermion loops. To have control over
these, we
consider the limit $N$ large with $g^2 N$ fixed.
Corrections from the scalar self-interactions will be treated classically,
so we assume that $\lambda$ is small.

In the large $N$ limit the only loop diagram that
contributes to the effective action for $\Phi$ is given
by Figure \ref{fig:oneloop} and
is proportional to $g^2 N$. All other diagrams are
suppressed by powers of $1/N$, see for example \cite{coleman}.
(Actually, at leading order there 
is also a two-loop diagram which
contributes to the cosmological constant).

To compute the effective action requires regularization
and
renormalization of the parameters in the theory.
We use dimensional regularization
and subtract using the $\overline{\hbox{MS}}$ scheme. In the
large $N$ limit there is mass and wavefunction renormalization for
$\Phi$, but none for either fermion. The only renormalization of
the coupling $g$ is due to the wavefunction renormalization of
$\Phi$. Denoting bare quantities with a subscript ``$0$''
and those without as the renormalized quantities,
one has
\beqa
\Phi^*_0 \Phi_0 &=& Z_{\Phi} \Phi^* \Phi \nonumber \\
g^2_0 Z_{\Phi} &=& g^2 \mu^{\epsilon} \nonumber \\
\lambda_0 Z^2_{\Phi} &=& \lambda \mu ^{\epsilon}\nonumber \\
m^2_{\Phi,0} Z_{\Phi} &=& m^2_{\Phi} + \delta m^2
\eeqa
with the renormalized quantities implicit functions
of the subtraction scale $\mu$ and $d=4 -\epsilon$. Then
\beq
Z_{\Phi} = 1 - \frac{g^2 N}{8 \pi^2} \left( \frac{2} {\epsilon} -\gamma_E
+ \ln (4 \pi) \right)
\eeq
\beq
\delta m^2 = -\frac{3 g^2 N}{4 \pi^2} m^2_{\psi} \left(\frac{2}{\epsilon}
- \gamma_E + \ln (4 \pi) \right)~.
\eeq
One finds the $\beta$ function
\beq
\beta(g) = - \frac{\epsilon}{2} g Z_{\Phi} ~.
\eeq
In the
limit $\epsilon \rightarrow 0$, one obtains the exact solution
\beq
\frac{1}{g^2(\mu)} = \frac{1}{g^2(\mu_0)} - \frac{N}{4 \pi^2}
\ln \left[\frac{\mu}{\mu_0}\right]~.
\eeq
As expected, the theory has a Landau pole.
One notes that since there is only wavefunction renormalization
of the coupling $g$, $ g^2 \Phi^* \Phi $
is independent of $\mu$.

In terms of the renormalized quantities, the full effective
action for $\Phi$ is
\beq
\int d^4 x {\cal L}_{\rm eff} = \int d^4 p \tilde{
\Phi}^*(-p) G(p^2) \tilde{\Phi} (p)- \int d^4 x V(|\Phi|)
\eeq
where
\beq
V= m^2_{\Phi}(\mu) \Phi^* \Phi + \frac{\lambda(\mu)}{4}(\Phi^* \Phi)^2 ~,
\eeq
\beqa
G(p^2) &=& g^2(\mu)
\left\{ \frac{p^2}{g^2(\mu)}
-\frac{Np^2 }{4 \pi^2 }  \ln \left[\frac{m_{\psi}}{\mu}\right]
+\frac{N }{4 \pi^2} \left(2 m^2_{\psi}-\frac{1}{2} p^2 \right)
\int^1 _0 dy \ln \left[ 1- y(1-y) \frac{p^2}{m^2_{\psi}}\right] \right. \nonumber \\
 & & \left.
-\frac{N}{4 \pi^2}m^2_{\psi}\left(1 -6\ln \left[\frac{m_{\psi}}{\mu}\right]
\right) \right\}
\label{exactsol}
\eeqa
and $\tilde{\Phi}(p)$ denotes the Fourier transform of $\Phi(x)$.
Since the effective action is exact, it should be independent
of $\mu$. Inspecting the above equation, there are several
sources for such a dependence. From the running of $g$, one
sees that the $\mu$-dependence of
the first two terms in $G$ cancel. In fact, together
they combine into $p^2/g^2(m_{\psi})$. The other
sources in $G$ of $\mu$-dependence is the over all
dependence on $g^2(\mu)$, and in the term in the last
line. The overall $\mu$ dependence proportional
to $g^2(\mu)$ cancels the wavefunction renormalization of
$\Phi$. The $\mu$ dependence of the last term is canceled
by the $\mu$ dependence of the $m^2_{\Phi}$ term
in the potential. Likewise, the other terms in the potential are
$\mu$ independent.
After shifting
$G(0)$ into the mass term, $\lambda$ and $m^2_{\Phi}$
run only by wavefunction renormalization.

The effective action has a cut beginning at the branch 
point $p^2=4 m^2_{\psi}$
corresponding to fermion pair production. For $p^2$ less
than this value we can power series expand in $p^2/m^2_{\psi}$, transform
to position space, then resum to get
\beq
{\cal L}_{\rm eff} = \Phi^* G(-\Box) \Phi - V(|\Phi|)~.
\label{exactmodel}
\eeq
Note that if the magnitude of $\Phi$ is frozen, 
then, $-\Box \rightarrow X$. 
The integral over the Feynman parameter $y$ can be done exactly for
$p^2 < 4 m^2_{\psi}$. One gets
\beq
G(-\Box)= g^2(\mu) \left(\frac{-\Box}{g^2(m_{\psi})}
+\frac{N}{4 \pi^2}
m^2_{\psi} f(-\Box/m^2_{\psi})\right)
\label{exactsol2}
\eeq
with
\beq
f(z) = \left(4- z\right) \left[-1 +\frac{\sqrt{4-z}}{\sqrt{z}}
\arctan (\frac{\sqrt{z}}{\sqrt{4-z}}) \right]~.
\eeq
Note that we have shifted $G(0)$ into the renormalized mass $m^2_{\Phi}$
so that in (\ref{exactsol2}) $G(0)=0$.

We now expand the effective action in powers of $\Box/m^2_{\psi}$.
For small $z$
\beq
f(z)= -\frac{z}{3} + \frac{z^2}{20} + \frac{z^3}{280} + \frac{z^4}{2520}
+ \cdots~.
\eeq
Then
\beq
G(-\Box) = g^2(\mu) \left( \frac{1}{g^2(m_{\psi})} - \frac{N}{12 \pi^2}
\right)(-\Box) + {\cal O}\left(\frac{\Box^2}{m^2_{\psi}} \right)~.
\eeq
Since the effective Lagrangian is independent of
$\mu $ (the running of $g^2$ is canceled by the wavefunction
renormalization) we are free to choose $\mu = m_{\psi}$.
Introducing
\beq
\gamma =\frac{g^2(m_{\psi})N}{12 \pi^2} -1
\eeq
the effective action for $\Phi$ at scales much below $m_{\psi} $
is
\beq
{\cal L}_{\rm eff} = -\gamma
 \partial ^{\mu} \Phi^* \partial _{\mu}
\Phi -V(\Phi^* \Phi)~.
\eeq
Terms suppressed by powers of $1/m^2_{\psi}$ have been dropped.

The important observation  at
this stage is that for large enough coupling $g^2(m_{\psi}) N$,
i.e. $\gamma>0$, the field $\Phi$ has a kinetic
term with the wrong sign. This is one of our main results.
Quantum corrections from the fermions have generated
a wrong-signed kinetic term.
However, the large coupling needed implies the fermion masses
are near the Landau pole. Nonetheless, our conclusions don't depend on the
use of dimensional regularization. 
In the Appendix we consider 
a general class of translation invariant regulators 
that 
cutoff the momentum on 
the order of $\Lambda$. For large enough bare coupling $g^2_0 N$
and by having the fermion masses $m_{\psi}$ of order $\Lambda$,
a wrong-signed kinetic term can be generated.

If $m_{\psi}$ is not treated 
as large, the model (\ref{exactmodel}) does lead to a minimum for
$P(X)$. However, the dynamics of $X$ (encoded in $P(X)$) 
cannot be separated from the fermion mass $m_{\psi}$, and 
hence the cutoff. We therefore take the limit $m_{\psi}$ large 
and use  
the results of section \ref{classicalphysics}, where it 
was shown that
classically integrating out scalars coupled to $\Phi$ will
always generate a correction to $P(X)$ that is monotonically
increasing. That will generate a
$P(X)$ of the form appearing in Figure \ref{fig:figgc1}, 
where the location of the minimum is now at a scale 
much below the ultraviolet cutoff. 

To illustrate how that can be
done, we choose the following 
simple model. We add a scalar $S$, also of charge $+1$, and only
consider mass mixing between $\Phi$ and $S$. Thus
to the Lagrange density (\ref{toymodelL}) we add 
\beq
\delta {\cal L}= \partial ^{\mu} S^* \partial_{\mu}
S - m^2_S S^* S - h^2 S^* \Phi -h^2  S \Phi^*
\eeq
where we have made a phase redefinition on 
$S$ so that $h$ is real and positive 
and further assume that $h < m_S$.
This model is not realistic, because other interactions consistent
with the symmetries of the theory have not been included. But
the
point here is just to illustrate that a toy model exists
which can generate a $P(X)$ will a local minimum.

Solving for $S$ at the classical level gives
\beq
S =- h^2 \frac{1}{\Box + m^2_S} \Phi
\eeq
leading to
\beq
{\cal L}_{\rm eff} = \Phi^* \left( \gamma \Box + h^4
\left[\frac{1}{\Box +m^2_S}
-\frac{1}{m^2_S} \right]
\right ) \Phi -V(\Phi^* \Phi)
\eeq
where we have shifted a mass squared term for $\Phi$ into the potential.
The kinetic term for $\Phi $ can be rewritten as
\beq
\Phi^* \left( \left[\gamma - \frac{h^4}{m^4_S}
\right]\Box + \frac{h^4}{m^4_S}
\frac{\Box^2}{\Box + m^2_S} \right) \Phi ~.
\eeq
Note that by choosing the coupling $g^2 N$ large enough and 
$\gamma > h^4/m^4_S$, one can maintain a wrong-signed
kinetic term for $\Phi$. Defining
\beq
\epsilon^{\prime} = \gamma - \frac{h^4}{m^4_S} >0~,
\eeq
the effective action for $\Phi$ is of the form
\beq
{\cal L}_{\rm eff} = \Phi^* F(-\Box) \Phi - V(\Phi ^* \Phi)
\eeq
where
\beq
F(z) =- \epsilon^{\prime} z + \frac{h^4}{m^4_S} \frac{z^2}{m^2_S - z}~.
\label{Fsol}
\eeq
For $z< m^2_S$ the second term is a monotonically increasing
function of $z$.

Next we look for solutions to the exact effective
action of the form
\beq
\Phi   = \sigma e^{i \phi}
\eeq
where $\phi = c t$. We will assume that in this
background $ \langle \sigma \rangle$ is nonzero
and constant, so that the $U(1)$ symmetry is broken.

For solutions with both $X$ and $\sigma $ constant, the
effective action is
\beq
{\cal L}_{\rm eff} =
\sigma^2 F(X) - m^2_{\Phi} \sigma^2 - \frac{\lambda}{4} \sigma^4~.
\eeq
We obtain the effective action for $X$ by solving for $\sigma$ and
integrating it out. One finds that
\beq
\sigma^2 =\frac{2}{\lambda} \left(F(X) - m^2_{\Phi} \right)
\eeq
and we assume that $-m^2_{\Phi}+ F(X)>0 $.
In terms of $v^2 = -2 m^2_{\Phi}/\lambda$, the vev of the potential at
$X=0$, one finds
\beq
P (X) =   v^2 F(X) +
\frac{F(X)^2}{ \lambda }~.
\label{peffquantum}
\eeq Then
\beq
P^{\prime} (X) = F^{\prime}(X) v^2 \left( 1+
 \frac{2 F(X)}{\lambda v^2}\right)= F^{\prime}(X)
\sigma^2~.
\eeq
Thus the sign and zeroes of $P^{\prime}$ are the same as for
$F^{\prime}$. In particular, with $\epsilon^{\prime} >0$, $P^{\prime}(X=0)<0$.

To determine the minimum of $P$ and study its
stability, it is convenient to assume $\epsilon^{\prime} \ll 1$.
This separates
the scale $\langle X \rangle $ from the scale
of $m_S$ or $v$, thus simplifying the analysis. Then
\beq
F(X)= -\epsilon^{\prime} X + \frac{h^4}{m^4_S}
\frac{X^2}{m^2_{S}} + {\cal O}\left(\frac{X^3}{m^4_S} \right)
\eeq
has a local minimum at
\beq
X = \frac{1}{2 \gamma} \epsilon^{\prime} m^2_{S} (1 + {\cal O}(\epsilon^{\prime}
))~.
\label{exsol}
\eeq
So with $\epsilon^{\prime } \ll \gamma \simeq h^4/m^4_S<1$,
$\langle X \rangle $ is much less than $m_S$.

\begin{figure}[t!]
  \centerline{\mbox{\epsfysize=6truecm
\hbox{\epsfbox{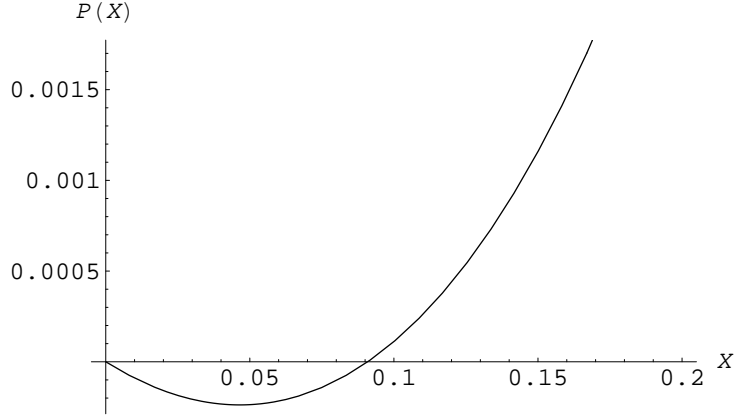}} } }
  {\caption[1]{$P(X)$ for $\epsilon^{\prime}=0.01$, $m_{S}=1$,
 $\lambda v^2/m^2_S=
1/9$ and $h^4/m^4_S = 1/10$. $P(X)$ is
in
units of $m^2_S v^2$.}
\label{fig:figPeff} }
\end{figure}

Inspecting the expression for $P$ in (\ref{peffquantum}),
one notes that the second term of ${\cal O}(F^2)$ is subdominant
to the first contribution in the limit $\epsilon^{\prime 2}
m^2_{S } \ll \gamma \lambda v^2 ~.
$ In this limit
the effective action for $X$ simplifies
to
\beq
P (X) \simeq F(X) v^2 ~,
\eeq
and the minimum for $X$ given by (\ref{exsol}) is a local
minimum of $P$. Using (\ref{Fsol}) and (\ref{peffquantum}),
a graph of $P(X)$ is given in
Figure \ref{fig:figPeff} for $\epsilon^{\prime}=0.01$,$~m_S=1$,
 $~\lambda v^2/m^2_S=
1/9$ and $h^4/m^4_S = 1/10$. $P(X)$ is given
in units of  $m^2_S v^2$.
The dark energy arising from the time-dependence of $\phi$ is of
order $\epsilon^{\prime 2} v^2 m^2_{S}/\gamma. $
Of course, in this toy model
there is in addition a contribution to the cosmological
constant from the potential which must be canceled.

\section{Stability Analysis}

In this section we 
study the stability of our theory. 
Since in an expanding universe $X$ is driven to $P^{\prime}(X)=0$ at late 
times \cite{nima1}, we 
focus
on the dynamics at that point. 

The model has two fluctuating degrees of freedom,
$\hat{\phi}$ and $\hat{\sigma}$ which are defined
by expanding about the background expectation value of $\Phi$,
\beq
\phi = ct + \hat{\phi}~~,~~\sigma  = \sigma_0 + \hat{\sigma}
\eeq
where
$\Phi = (\sigma_0+ \hat{\sigma} ) e^{ict +i \hat{\phi} } $.
It is convenient to introduce $c_{\mu} \equiv c \delta_{\mu 0} $
and recall that
$X = c_{\mu} c^{\mu}$. As discussed in the previous
section, $c_{\mu}$ and $\sigma_0$ are given by solving
the classical equations of motion. We assume the parameters
of the model are such that $\sigma_0 \neq 0$ and 
restrict our attention 
to terms with up to 
two-time derivatives acting on the fields. 

We begin by studying the dynamics of $\hat{\sigma}$. About
the minimum of the potential it will have a positive mass
squared. Its dependence on $X$ and the parameters of
the model is not needed here. For the stability
analysis, it is sufficient to determine terms in the
effective action involving only two
$\hat{\sigma}$ fields. Defining
$D_{\mu} \equiv \partial_{\mu} + i c_{\mu}$, the action to quadratic
order in $\hat{\sigma}$ is given by
\beq
{\cal L }_{\rm eff}(\sigma) = 
\sigma F(-D^2) \sigma - \frac{\lambda}{4}(\sigma^2 - v^2)^2 ~.
\eeq
With $D^2= \partial^2 + 2i c \cdot \partial -c ^2$, expanding the
action about $D^2=-c^2$ to the four-spatial derivative and 
two-time derivative levels gives, for
the terms involving derivatives only,
\beqa
{\cal L }_{\rm eff}(\sigma)&=&
 F^{\prime \prime}(c^2) \hat{\sigma}
\left[-2(c \cdot \partial )^2 + \frac{1}{2} \Box^2 \right]\hat{\sigma}
+ \cdots ~.\eeqa
We note that if $F^{\prime \prime}(c^2) >0$ then
$\hat{\sigma}$ has a healthy two-time derivative
term.

The effective
Lagrangian for $\hat{\sigma}$ has no
two-gradient terms. The leading gradient terms
then come from the second term involving
four spatial derivatives and is
\beq
\frac{1}{2} F^{\prime \prime}(c^2) \hat{\sigma} \nabla^4 \hat{\sigma}~.
\eeq
Since for stability of the kinetic energy (i.e. two time-derivative
term) we require that
$F^{\prime \prime}>0$, the four-gradient
term has the wrong sign. The dispersion
relation for $\hat{\sigma}$ is then $
4 c^2 F^{\prime \prime}(c^2) E^2 = m^2_{\sigma} - F^{\prime \prime}(c^2) k^4$.
The instability only occurs at short wavelengths of order
$k^4 \simeq m^2_{\sigma}/F^{\prime \prime}(c^2)$.

In this model the stability of the kinetic energy
and the gradient energy of $\hat{\sigma}$ place opposing
requirements on the
curvature of $F$.
Fortunately
it is not difficult to enlarge the model generating
the correct sign for $\hat{\sigma}$'s
two-gradient term.
For example, consider a massive real scalar $S^{\prime}$ neutral
under the $U(1)$ symmetry and the following Lagrangian
\begin{equation}
{\cal L}_S = -\frac12 S^{\prime} \Box  S^{\prime}
 - \frac12 m_{S^{\prime}} ^2 S^{\prime 2} + g^{\prime} S^{\prime}
\Phi^* \Phi ~.
\end{equation}
By integrating out $S^{\prime}$ at the tree-level we obtain the following
effective Lagrangian
\begin{equation}
\delta {\cal  L}_{\rm eff} =
\frac{g^{\prime 2}}{2 m^2_{S^{\prime}} }\Phi^* \Phi
 \frac{1}{1+\Box /m^2_{S^{\prime} }} \Phi^* \Phi
\label{sigmaop}
\end{equation}
which gives a healthy kinetic term to the radial mode at the
two-derivative level,
\begin{equation}
\delta {\cal L}_{\rm eff} =
-\frac{g^{\prime 2}}{2m^4_{S^{\prime}}}
\Phi^* \Phi \Box (\Phi \Phi^*)  \rightarrow
-\frac{2 g ^{\prime 2} \sigma^2_0}{m^4_{S^{\prime}}}
 \hat{\sigma} \Box \hat{\sigma} ~.
\end{equation}
We assume that the coefficient of this contribution 
is large enough 
to stabilize $\hat{\sigma}$. 
Note that since $S^{\prime}$ is neutral, the above modification
to the effective action for $\Phi$ is independent of
the Goldstone boson.

Next we turn to the stability condition for the Goldstone field
$\hat{\phi}$. As before, we investigate the dynamics 
at $P^{\prime}=F^{\prime}=0$. Since we have 
stabilized $\hat{\sigma}$ in the preceding discussion, we 
ignore its fluctuations in discussing the stability 
of the Goldstone boson 
\footnote{Properly integrating out
${\sigma}$ does not change this conclusion. From
solving the classical equations of motion at 
the point $P^{\prime}(c^2)=0$, one finds an
additional contribution to $\sigma$ involving spatial
gradients that is of the form
$\delta {\sigma} \propto F^{\prime \prime} \sigma_0 \left[ e^{-i \hat{\phi}}
\nabla^4 e^{i \hat{\phi}} +{\rm h.c.} \right]$. The point is
that since $\delta {\sigma} \sim \nabla^4 \hat{\phi}$, inserting it back
into the action generates terms for $\hat{\phi}$ involving
six spatial derivatives.}.  
Then the effective action 
for $\hat{\phi}$ is 
\beq
{\cal L}_{\rm eff} = \sigma_0 ^2 e^{-i \hat{\phi}} F(-D^2) e^{i \hat{\phi}}~.
\eeq
The part of the effective action involving two $\hat{\phi}$ fields is
\beqa
{\cal L}_{\rm eff} &=& \frac{1}{2}F^{\prime \prime}(c^2)
\left[ \hat{\phi} \Box^2 \hat{\phi} - 4 \hat{\phi} (c \cdot
\partial )^2 \hat{\phi} \right] +\cdots ~.
\eeqa

Here 
$\hat{\phi}$ has the correct sign for
its two-time derivative kinetic term if 
$P^{\prime \prime} = F^{\prime \prime}
\sigma_0^2 >0$. As in \cite{nima1}, its two-derivative
gradient term vanishes. Unfortunately, $\hat{\phi}$ has the wrong
sign for its four-derivative gradient term since $P^{\prime \prime}>0$.
This leads to the dispersion relation (neglecting terms involving 
$E^4$ and $k^2 E^2$) 
$ E^2 \simeq -k^4/ 4 c^2$, 
which has instabilities at small $k$.

These spatial instabilities can be removed at the expense of adding higher
dimensional operators.
One possibility is to add
\begin{equation}
{\cal L}_{\hat{\phi}} =  
+\frac{\lambda^{\prime}}{2}(\Phi^* \Box \Phi)^2 + {\rm h.c.}~,  
\end{equation} 
with $\lambda^{\prime}>0$. 
This operator contributes  $\delta P(X) =\lambda^{\prime} X^2 \sigma^4_0$, 
shifting the 
minimum of $P(X)$ and $\sigma_0$, 
but not affecting $P^{\prime}(X=0)<0$. 
Expanding this operator (about the new minimum) 
to quadratic order in small fluctuations 
for $\hat{\phi}$, 
one finds
\begin{equation} 
\delta {\cal L}_{\hat{\phi}} = \lambda^{\prime} \sigma^4_0 \left(
-\hat{\phi} \Box^2 \hat{\phi} - 4\hat{\phi} 
(c \cdot \partial)^2  \hat{\phi}  \right) 
\label{gbfix}
\end{equation}
where we have dropped terms with more derivatives.
Note that with $\lambda^{\prime}>0$, this operator
gives the correct sign to terms involving four spatial derivatives
and to terms involving two time derivatives acting on $\hat{\phi}$.
Thus when added to the Lagrangian with
large enough coefficient, it
can remove the instability in the four-gradient term without affecting the
requirement that the kinetic energy term involving two time derivatives 
is positive or the  
behavior of $P^{\prime}(X=0)$. 
Although the operator in (\ref{gbfix})
is not renormalizable, it may be possible
to generate it by integrating out heavy degrees of freedom.

\section{Conclusions}

The Higgs phase of gravity or ghost condensation is an interesting
proposal for the dark energy. There the dark energy arises
dynamically from a time-dependent scalar field $\phi$ that
spontaneously breaks Lorentz invariance. The effective action for the
$\phi$ contains only derivatives of the field so it is natural to
hypothesize that  $\phi$ is the Goldstone boson resulting from a
spontaneously broken $U(1)$ symmetry. It is conventional to introduce
the notation $X=(\partial_{\mu}\phi)^2$ and use $P(X)$ to 
denote the part of the 
Lagrangian that contains single derivatives on $\phi$.
Usually the evolution of the universe redshifts away the energy
density stored in the time dependence of $\phi$. However, if there is
a value of $X$ where $P^{\prime}(X)=0$ then Lorentz symmetry is
broken asymptotically as $t \rightarrow \infty$ and the time
dependent scalar field is a candidate for the dark energy. In Figures
\ref{fig:figgc1} and \ref{fig:figgc2} examples of such Lagrange
densities are shown. In the first case $P^{\prime}(0)<0$ which
corresponds to a wrong sign kinetic term for $\phi$.
 
In this paper we studied the possibility that the Higgs phase of
gravity is the low energy limit of an underlying theory that is
renormalizable, has standard kinetic terms for its fields and
spontaneously breaks a $U(1)$ symmetry. In particular we are
interested in whether a $P(X)$ of the form shown in  Figure 
\ref{fig:figgc1} or Figure \ref{fig:figgc2} can be generated by
integrating out the heavy degrees of freedom in such models.
 
Our results show that in a wide class of
theories involving only scalars, the spontaneous
breaking a global $U(1)$ symmetry leads, at the classical level, to
an
effective action $P(X)$ for the Goldstone boson that is
always monotonically increasing. While such an effective action can
have
solutions that break Lorentz invariance,
those solutions are not relevant at late times
in an expanding FRW universe.
 
We examined a model where the field whose vev spontaneously
breaks the global $U(1)$ symmetry is coupled to $2N$ massive
fermions and another complex scalar. 
In the large $N$ limit the effective action
for the scalar field that directly couples to the fermions 
can be computed exactly. At large
enough $g^2 N$, an effective action
of the form displayed in Figure \ref{fig:figgc1} is obtained.
 
Unfortunately this model is not a satisfactory high energy theory for
two reasons. The theories we consider are renormalizable but 
still require regularization via an ultraviolet cutoff. Unlike 
in conventional theories, here we find
that the fermion masses cannot be taken to be small
compared with the cutoff \footnote{We considered 
two different regulators, a general 
translation invariant momentum cutoff satisfying 
very reasonable assumptions, and 
dimensional regularization. Our conclusions do not depend 
on the explicit choice of regulator.}.
Secondly there is a spatial instability for the Goldstone boson.
By extending the model considered in this paper, 
it may be possible 
to overcome the latter difficulty. 
Our work is progress towards the goal of finding
high energy completions for low energy effective theories with ghost
condensation.

\vspace*{.5cm}
\noindent
{\bf Acknowledgements:}
This work was supported by grants from the Department of Energy under
DE-FG02-90ER40542 at IAS and DE-FG03-92ER40701 at Caltech.
I.L. would like to thank
the theory group at Caltech for hospitality
during the completion of this work.

\appendix
\section*{Appendix}

Since in the model of section \ref{toymodel} the fermion
masses were of order the Landau pole, one might worry
that the conclusions of that section depend on the use
of dimensional regularization.
In this Appendix we compute the effective action for
$\Phi$ using the Lagrangian
in (\ref{toymodelL}) but regulate the
theory with a momentum cutoff rather than with dimensional regularization.
We shall see that
for a range of values for $m_{\psi}$ of order
$ \Lambda$ and
large enough bare coupling $g^2_0 N$,
quantum corrections to the effective action
can generate a wrong-signed kinetic term for $\Phi$.

\begin{figure}[t!]
  \centerline{\mbox{\epsfysize=6truecm
\hbox{\epsfbox{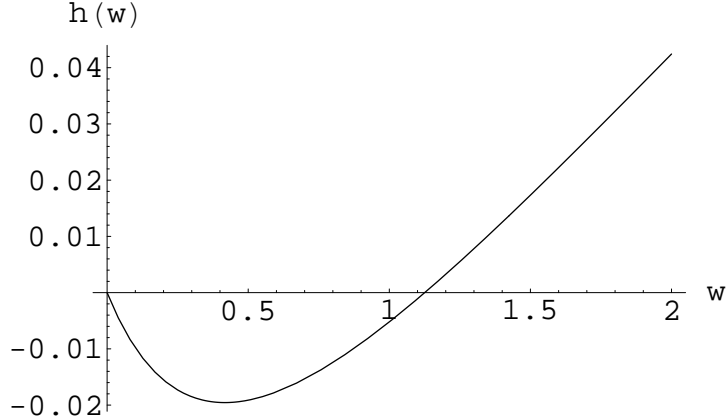}} } }
  {\caption[1]{$h(w)$. Note that $h<0$ for a range of 
$w = \Lambda^2/m^2_{\psi}$. The growth for large 
$w$ is logarithmic. }
\label{fig:app} }
\end{figure}

As before, at large $N$ the Feynman diagram in Figure \ref{fig:oneloop}
is the exact quantum correction to the effective
action for $\Phi$. To regulate the diagram we first 
Wick rotate both the external and internal momenta into 
Euclidean space. At the end of the computation, we Wick 
rotate the external momentum back to Minkowski signature. 
To preserve translation invariance, we regulate the Euclidean 
space propagators by modifying them accordingly \cite{polchinski}, 
\beq 
 \frac{1}{(k+p)^2 + m^2_{\psi}} \rightarrow 
\frac{1}{(k+p)^2+m^2_{\psi}} {\cal K}\left[(k+p)^2 /\Lambda^2\right] 
\eeq 
where ${\cal K}[x] \rightarrow 0$ for large $x$. 
To begin with, the regulator we consider is 
\beq 
{\cal K}[q^2] =
e^{-q^2/\Lambda^2} ~.
\label{expreg}
\eeq
In (Euclidean) position space this is equivalent to introducing a 
term $e^{-\Box/\Lambda^2}$ into the kernel. 
To extract the $p^2$ coefficient, we expand all 
terms to that order. Focusing on the $p^2$ terms only gives 
\beq 
{\cal L}_{\rm eff} = \Phi^* F(-\Box) \Phi
\label{AppLeff}
\eeq
(now $\Box$ is in Minkowski space) where 
\beq
F(z) = z + \frac{g^2_0 N}{4 \pi^2} z h(w)
\label{Apptreeresult}
\eeq 
with 
\beqa 
h(w) &=& 
\int ^{\infty} _0 dx
e^{-2 x/w} \frac{x}{(x+1)^2} \left( 
(1-x) \left[- \frac{1}{w} - \frac{1}{x+1} + \frac{x}{(x+1)^2} 
+ \frac{1}{2} \frac{x}{w^2} + \frac{x}{x+1} \frac{1}{w} \right]
\right. \nonumber \\ 
& & \left.  
+ \frac{1}{2} \frac{x}{x+1} + \frac{1}{2} \frac{x}{w} \right) 
\label{Appresult2}
\eeqa
and $ w \equiv \Lambda^2/m^2_{\psi}$. 
The first term appearing in (\ref{Apptreeresult}) is the tree result. 
By inspecting (\ref{Appresult2}) in the limit $w \ll 1$ one can see
without much work that $h(w)<0$ for small $w$.  
A plot of $h(w)$ is shown in Figure \ref{fig:app} for a larger 
range of $w$. Note that for
$0 \leq \Lambda^2 /m^2_{\psi} \leq 1.1$  the function $h$ is negative.

Recalling the result from section \ref{classicalphysics}
that
\beq
P^{\prime}(X)= \frac{\partial F}{\partial X}
\eeq
evaluated at $\sigma(X)$,
\beq
P^{\prime}(X=0)= \sigma^2
\left[ 1 + \frac{g^2_0 N}{4 \pi^2} h\left(\frac{\Lambda^2}{m^2_{\psi}}\right)
\right]~.
\label{Pprime0cutoff}
\eeq
Inspecting the result (\ref{Pprime0cutoff}) for $P^{\prime}(X=0)$,
we see that since there is a range for 
$\Lambda^2/m^2_{\psi}$ for which $h<0$, 
then for each $\Lambda/m_{\psi}$ 
there exists a critical coupling $g^2_{\rm crit} N$ such
that for $g^2_0 N> g^2_{\rm crit} N$, $P^{\prime}(X=0)<0$.

Next we take the limit that both $\Lambda$ and $m_{\psi}$ are
both very large (say of the order the Grand Unification scale),
but holding $w=\Lambda^2/m^2_{\psi}$ fixed at a value
for which $h(w)<0$. We also choose $g^2_0N > g^2_{\rm crit}N$. 
Then the effective theory for $\Phi$
far below the cutoff is
\beq
{\cal L}_{\rm eff} = -\gamma \partial ^{\mu} \Phi^*
\partial_{\mu} \Phi -V(\Phi^*\Phi)~,
\eeq
where $\gamma=-1-g^2_0 N h(\Lambda^2/m^2_{\psi})/4 \pi^2>0$.
Higher order terms
are suppressed by $p^2/\Lambda^2$ and are irrelevant.

The above results were obtained using the exponential regulator 
(\ref{expreg}). The conclusion that $h<0$ for $\Lambda^2 \ll m^2_{\psi}$ 
is true, however, for a general ${\cal K}$. To see that, note 
that in the limit $\Lambda^2 \ll m^2_{\psi}$, the $p$-dependence 
of the two-point function is dominated by  
the factor ${\cal K}[(k+p)^2/\Lambda^2]$ in the loop integral. The dominant 
contribution to $h$ in this limit is then  
\beqa 
h\left(\frac{\Lambda^2}{m^2_{\psi}}\right) &=& 
\frac{m^2_{\psi}}{\Lambda^2}  
\int ^{\infty}_0 dx ~x {\cal K}[x]\left( 
{\cal K}^{\prime}[x] + \frac{1}{2} x {\cal K }^{\prime \prime}[x] \right) 
\nonumber \\ 
& =& - \frac{m^2_{\psi}}{2 \Lambda^2} \int ^{\infty}_0 
dx~x^2 {\cal K}^{\prime 2 } <0 ~.
\eeqa
To arrive at the second line 
we have integrated by parts and 
assumed 
that $x^2 {\cal K} {\cal K}^{\prime}(x) \rightarrow 0$ as 
$x \rightarrow \infty$. Thus $h<0$ is true quite generally. 

We have shown
that in a theory regulated by a 
general function ${\cal K}(q^2/\Lambda^2)$, a negative kinetic
term for $\Phi$ can be
generated by choosing a large enough bare coupling $g^2_0 N$ and
a value for $m_{\psi}$ of order $\Lambda$ such
that $h(z)<0.$ 
For large $\Lambda$, higher dimension operators
are suppressed by $\Lambda^2$. As discussed in
section \ref{toymodel}, introducing other scalars
coupled to $\Phi$ can give rise to a minimum for 
$P(X)$ by 
generating, at the classical level, a correction to
$P(X)$ that is monotonically increasing.


\end{document}